%% file: CrSi2_arXiv.tex
\documentclass[%
superscriptaddress,
 preprint,
 amsmath,amssymb,
prb,
]{revtex4-2}

\usepackage{dcolumn}
\input{revpackage.tex}

\renewcommand{\addb}[1]{#1}
\renewcommand{\addr}[1]{#1}

\begin{document}


\title{Chiral phonons in sixfold chiral \ce{CrSi2}:\\Raman spectroscopy and first-principles calculations}

\author{Gakuto~Kusuno}
\email[]{kusuno.g.2b7e@m.isct.ac.jp}
\affiliation{%
Department of Physics, Institute of Science Tokyo, Meguro-ku, Tokyo 152-8551, Japan%
}%
\author{Shingo~Kisanuki}
\affiliation{%
Department of Physics and Electronics, Osaka Metropolitan University, Sakai, Osaka 599-8531, Japan%
}%
\author{Yusuke~Kousaka}
\affiliation{%
Department of Physics and Electronics, Osaka Metropolitan University, Sakai, Osaka 599-8531, Japan%
}%
\author{Yoshihiko~Togawa}
\affiliation{%
Department of Physics and Electronics, Osaka Metropolitan University, Sakai, Osaka 599-8531, Japan%
}%
\affiliation{%
Quantum Research Center for Chirality, Institute for Molecular Science, Okazaki, Aichi 444-8585, Japan%
}%
\author{Takuya~Satoh}
\email[]{satoh@phys.sci.isct.ac.jp}
\affiliation{%
Department of Physics, Institute of Science Tokyo, Meguro-ku, Tokyo 152-8551, Japan%
}%
\affiliation{%
Quantum Research Center for Chirality, Institute for Molecular Science, Okazaki, Aichi 444-8585, Japan%
}%

\date{\today}

\begin{abstract}
Chiral phonons have been identified in several chiral crystals, primarily in those with trigonal symmetry 
and threefold screw axes. In this study, chiral phonons in \ce{CrSi2}, 
a chiral crystal with a sixfold helical structure, were investigated. 
Circularly polarized Raman spectroscopy revealed
a subtle splitting of doubly degenerate $E_2$ phonon modes between cross-circular polarization configurations. 
These observations, supported by first-principles phonon calculations, indicate the presence of 
chiral phonons in \ce{CrSi2}, expanding the scope of materials that exhibit chiral vibrational modes 
beyond the conventional trigonal class.
\end{abstract}

\maketitle


\section{\label{sec:intro}Introduction}
In recent years, rotational atomic motions in crystals have attracted significant attention in the field of condensed-matter physics.
Phonons endowed with angular momentum arising from such rotations have been theoretically predicted and experimentally observed in connection with diverse phenomena~\cite{wang2024}, including
the phonon Zeeman effect~\cite{juraschek2019a,juraschek2022a,luo2023b,lujan2024,che2025}, 
the phonon Hall effect~\cite{grissonnanche2020,li2020a}, 
and their topological characteristics~\cite{zhang2023d,zhang2025z}.
Although these phonons have been investigated extensively in two-dimensional materials~\cite{zhang2015,zhu2018a}, chiral crystals offer distinct opportunities,
such as mechanical rotation along an axis parallel to the temperature gradient~\cite{hamada2018} and
chirality-induced selectivity of phonon polarizations~\cite{chen2022,ohe2024}.
In particular, phonons that carry finite angular momentum and propagate with a finite wavevector along the rotation axis are referred to as ``chiral phonons'' \addb{that satisfy the definition of true chirality}~\cite{barron1986,juraschek2025}.

Observation of chiral phonons in chiral materials has been achieved using circularly polarized Raman spectroscopy in 
$\alpha$-\ce{HgS}~\cite{ishito2023a}, \ce{Te}~\cite{ishito2023c}, and $\alpha$-quartz~\cite{oishi2024}
and also by resonant inelastic X-ray scattering in $\alpha$-quartz~\cite{ueda2023}.
These materials share the same threefold helical structure, with space groups limited to \addb{$P3_221$ (left-handed)} and \addb{$P3_121$ (right-handed)}.
In such chiral materials, the energy splitting of chiral phonons with opposite angular momenta in reciprocal space makes them particularly suitable for exciting chiral phonons~\cite{ishizuka2025z}.
However, experimental evidence of chiral phonons in other types of chiral structure has been lacking.

Sixfold helical structures are of particular interest because of their similarities to their trigonal counterparts.
For instance, certain transition metal disilicides ($M\mathrm{Si_2}$, $M=\ce{V},\pad \ce{Nb},\pad \ce{Ta}$) crystallize in sixfold helical structures~\cite{chaix-pluchery1997,fujio2007}.
These materials exhibit chirality-related phenomena, such as chirality-induced spin selectivity with long-range propagation~\cite{shiota2021,shishido2021,roy2022}, as well as characteristic electronic band structures~\cite{tsutsumi2013,onuki2014,zhang2023c,mahraj2025}.

While these chiral disilicides are metallic, less conductive materials are more suitable for investigation by Raman scattering.
In this study, the focus was on the semiconductor disilicide \ce{CrSi2}, which possesses a sixfold helical axis with \addb{left-handed $\SGl$ ({\lCS})} and \addb{right-handed $\SGr$ ({\rCS})} enantiomorphs.
Circularly polarized Raman spectroscopy was applied to \ce{CrSi2} crystals.
The results demonstrate that the direction of energy splitting in doubly degenerate phonon modes can be predicted based on the phonon \addb{crystal-} \addb{(or pseudo-)}angular momentum (\addb{CAM}), analogous to the case of trigonal chiral materials.
Moreover, it was found that each split peak corresponded to a pair of chiral phonons carrying opposite angular momenta and opposite \addb{CAM}.

\section{Methods}
\subsection{Sample preparation}
\begin{figure}
  \centering
  \includegraphics[width=\linewidth]{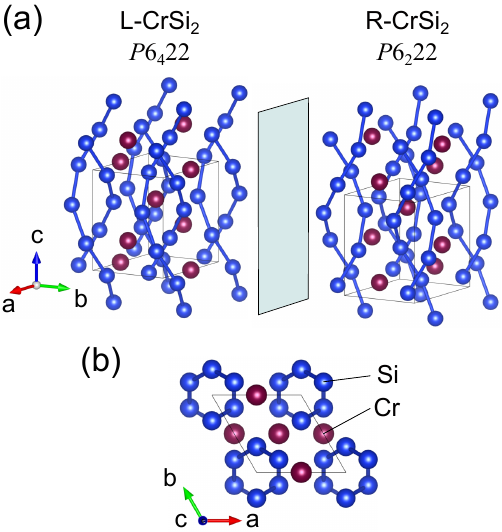}
  \caption{\label{fig:structure}Crystal structures of \ce{CrSi2}:
  (a) L- and {\rCS}, which are mirror images of each other, and
  (b) sixfold crystal structure viewed along the $c$-axis.}
\end{figure}

\ce{CrSi2} is a nonmagnetic semiconductor with a bandgap of \SI{0.35}{eV}~\cite{bost1988,mattheiss1991,bellani1992,chen2018}.
It crystallizes into a C40 structure with a sixfold screw axis~\cite{mattheiss1992}.
Its chirality arises from double helical chains formed by \ce{Si}--\ce{Si} and \ce{Si}--\ce{Cr} bonds.
The two enantiomeric structures are related to each other by either spatial inversion or mirror symmetry, as illustrated in Fig.~\subref{fig:structure}{a}.

Crystals of \ce{CrSi2} were synthesized using two methods: the laser-diode-heated floating zone (LDFZ) method~\cite{kousaka2023} and the chemical vapor transport (CVT) method~\cite{binnewies2012,szczech2007}.
In the LDFZ method, the handedness of the synthesized crystals can be controlled by using a seed rod with a predetermined handedness.
In contrast, the CVT method does not provide selective control over crystal handedness.

Two enantiomeric crystals of \ce{CrSi2} were synthesized using the LDFZ method.
The first crystal, grown from a left-handed seed, was identified as left-handed ({\lCS}) by absolute structure analysis using single-crystal X-ray diffraction (XRD).
The Flack parameter~\cite{flack1983} was determined to be \num{-0.114 \pm 0.134} (with the uncertainty corresponding to one standard deviation: $\sigma$), confirming its handedness.
The uniform handedness throughout the entire crystal, approximately \SI{3}{mm} in size, was further verified by circularly polarized resonant X-ray diffraction~\cite{tanaka2008}.
The second crystal was grown using a right-handed seed rod and identified as right-handed ({\rCS}) based on XRD measurements, with a Flack parameter of \num{0.098 \pm 0.116}.

Additional crystals were synthesized by the CVT method in a temperature gradient using iodine as the transport agent.
The CVT-grown crystals typically exhibited hexagonal shapes with sizes of approximately \SI{200}{\um}.
Sample~\#1 was identified as left-handed by XRD, with a Flack parameter of \num{-0.002 \pm 0.142}.
For Sample~\#2, however, the large uncertainty ($\sigma > 0.25$) prevented a reliable determination of its handedness.

\subsection{\label{sec:Raman}Circularly polarized Raman spectroscopy}
All Raman measurements were conducted at room temperature (\SI{295}{K}) using an excitation wavelength of \SI{785}{nm}
in the backscattering geometry perpendicular to the $c$-plane of the crystal [Fig.~\subref{fig:structure}{b}].
The Raman measurement system was \addr{custom-built and equipped with a spectrometer (Princeton Instruments, SpectraPro 2500i) and a CCD array sensor (Hamamatsu Photonics, S11501-1007S).}
\addb{Two} polarizers and a quarter-wave plate \addb{allow} full control of the incident and scattered light polarizations.
For example, RL denotes the right-circularly polarized incident light and left-circularly polarized scattered light (see the inset of Fig.~\ref{fig:mode_assignment}).
The cross-circular polarization \addb{(LR and RL)} configurations are particularly useful for detecting chiral phonons, because the phonon \addb{CAM} excited in these configurations is well separated.
Because of the energy splitting of doubly degenerate phonon modes with different \addb{CAM} at finite wavevectors along the $\Gamma$--A (helical chain) direction of the Brillouin zone,
chiral phonons can be detected as small differences in Raman shift between the cross-circular \addb{polarization} configurations~\cite{ishito2023a,ishito2023c,zhang2023d,oishi2024}.

The unit cell of \ce{CrSi2} contains three Cr atoms and six Si atoms, yielding 27 phonon modes.
Among these, the Raman-active modes are classified as $\Gamma=A_1 + 4E_1 + 4E_2$.
The doubly degenerate $E_1$ modes can be observed under oblique incidence relative to the $c$-axis of the crystal.
However, because the Raman measurements were performed under normal incidence along the $c$-axis, the $E_1$ modes were suppressed.
The other doubly degenerate modes, $E_2$, were observed in the cross-circular polarization configurations,
whereas the non-degenerate $A_1$ modes appeared in the parallel-circular polarization \addb{(LL and RR)} configurations.
\addr{The rotational motions of the four $E_2$ phonon modes are shown in Supplementary Movie~\cite{supple}.}


\subsection{First-principles calculations}
First-principles phonon calculations were performed using density functional perturbation theory, as implemented in the ABINIT software package~\cite{gonze2020,gonze1997a,gonze1997}, to verify the experimental results.
All calculations employed the local density approximation for the exchange-correlation functional formulated by Perdew and Wang~\cite{perdew1992}.
The projector augmented-wave method~\cite{blochl1994,torrent2008} was used for pseudopotentials, and a plane-wave energy cutoff of \SI{25}{Ha} was applied.
The Brillouin zone was sampled using a $4\times4\times4$ Monkhorst--Pack $k$-point grid~\cite{monkhorst1976}, shifted to include the $\Gamma$ point.
Phonon dispersion relations were calculated on a $4\times4\times4$ $q$-point grid.

\section{\label{sec:result}Experimental results and discussion}
\subsection{Raman measurements of LDFZ-grown crystals}
\begin{figure}
  \centering
  \includegraphics[width=\linewidth]{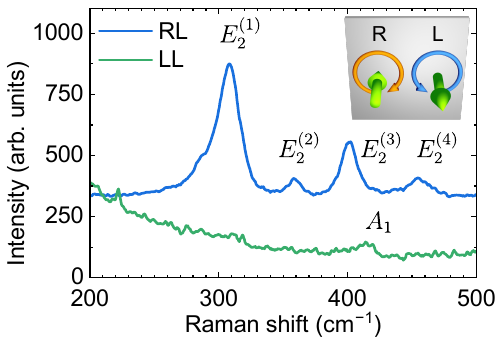}
  \caption{\label{fig:mode_assignment}\addr{Stokes Raman spectra and} \addb{p}honon mode assignments for the {\lCS} crystal:
  \addb{vertical offsets have been added for clarity.}
  \addb{T}he inset shows the RL configuration. Here, R and L are defined as the direction of polarization rotation in the sample plane, irrespective of the propagation direction.}
\end{figure}

\addr{Figure~\ref{fig:mode_assignment} shows the Stokes Raman spectra of the {\lCS} crystal grown by the LDFZ method, together with the corresponding phonon mode assignments.}
The $E_2$ and $A_1$ modes were observed in the cross- and parallel-circular polarization \addb{(RL and LL)} configurations, respectively.
The observed Raman shifts\addr{,} \addr{obtained by fitting the spectra with Voigt functions,} \addb{and the mode assignments} were consistent with our first-principles calculations \addb{and Raman selection rules (see Appendix~\ref{app:selection_rule})} as well as with reported values~\cite{hermet2015}, as summarized in Table~\ref{tab:raman_shift}.

\addr{The $E_2^{(1)}$ mode can be fitted by two Voigt functions.
Under the ideal backscattering geometry along the $c$-axis, the $E_1$ modes are symmetry-forbidden. 
However, slight misalignment from the ideal geometry can weakly relax the selection rules, leading to small contributions from $E_1$ modes. 
We attribute the observed two-peak structure of the $E_2^{(1)}$ mode to such weak $E_1$-mode contributions, whose frequencies are close to that of the $E_2^{(1)}$ mode (see Table~\ref{tab:raman_shift}).
These contributions are extrinsic and negligible compared with the intrinsic chiral splitting discussed in this work, and are therefore excluded from the subsequent peak-fitting analyses.}

\begin{table}
  \caption{\label{tab:raman_shift}Raman shifts (\unit{cm^{-1}}) of observed and calculated phonon modes.}
  \begin{ruledtabular}
    \begin{tabular}{cccccccc}
      & {$E_1^{(3)}$} & {$E_2^{(1)}$} & {$E_1^{(4)}$} & {$E_2^{(2)}$} & {$E_2^{(3)}$} & {$A_1$} & {$E_2^{(4)}$} \\
      \midrule
      Exp. & \addr{292} & \addb{308} & {-} & 359 & 401 & 412 & 456 \\
      Calc. & \addr{316} & 339 & \addr{370} & 377 & 423 & 436 & 484 \\
      Ref.~\cite{hermet2015} (Exp.) & \addr{301} & {307--311} & \addr{357} & \addb{354} & 399 & 413 & {-} \\
      Ref.~\cite{hermet2015} (Calc.) & \addr{282} & 311 & \addr{336} & 338 & 388 & 403 & 448 \\
    \end{tabular}
  \end{ruledtabular}
\end{table}

\begin{figure}
  \centering 
  \includegraphics[width=0.8\linewidth]{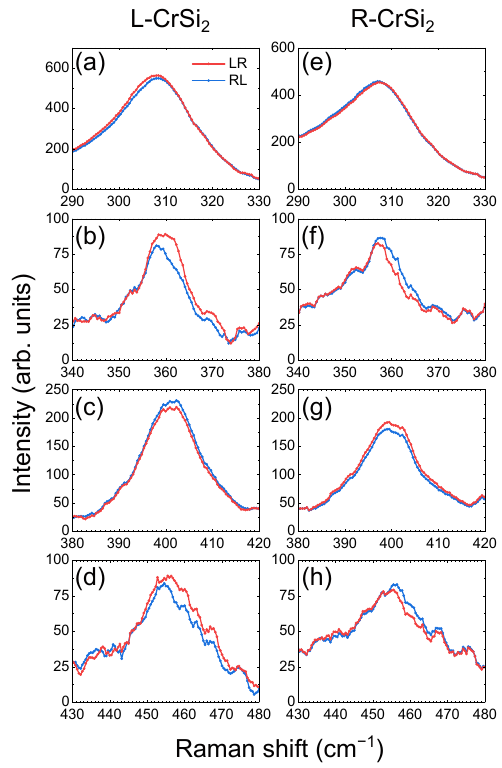}
  \caption{Stokes spectra of the four $E_2$ modes of (a--d) L- and (e--h) {\rCS} in cross-circular polarization configurations:
  (a), (e) $E_2^{(1)}$, (b), (f) $E_2^{(2)}$, (c), (g) $E_2^{(3)}$, and (d), (h) $E_2^{(4)}$ peaks,
  showing polarization-dependent energy splittings.}
  \label{fig:fz_spectra}
\end{figure}

Stokes spectra of the {\lCS} crystal obtained in the cross-circular polarization \addb{(LR and RL)} configurations are shown in Figs.~\subref{fig:fz_spectra}{a}--\subref{fig:fz_spectra}{d}.
All four $E_2$ phonon modes exhibited small but discernible differences in Raman shift between the \addb{LR} and \addb{RL} spectra.
The RL configuration exhibited higher Raman shifts than the LR configuration for the $E_2^{(1)}$ and $E_2^{(3)}$ modes, whereas the opposite trend was observed for the $E_2^{(2)}$ and $E_2^{(4)}$ modes.
\addb{Moreover}, the Stokes spectra of the {\rCS} crystal exhibited energy splittings in the opposite direction [Figs.~\subref{fig:fz_spectra}{e}--\subref{fig:fz_spectra}{h}], which is consistent with the opposite crystal handedness.

\addr{To extract the splitting widths quantitatively, all Raman peaks were fitted using a Voigt profile, given by the convolution of a Gaussian and a Lorentzian function. 
The Gaussian width was fixed to \SI{2.7}{cm^{-1}} to represent the instrumental response, while the Lorentzian component accounts for the intrinsic phonon linewidth. 
The uncertainties of the splitting widths listed in Tables~\ref{tab:compare_exp_tho} and \ref{tab:compare_sample1_2} correspond to the standard errors obtained from these Voigt-function fits. 
The observed splittings are larger than the fitting uncertainties and are well resolved within the signal-to-noise ratio of the measured spectra, demonstrating that the splittings are statistically significant despite their small magnitude.}

\addr{The observed splitting magnitudes ($\leq \SI{1}{cm^{-1}}$) in \ce{CrSi2} are comparable to those reported in trigonal chiral crystals such as $\alpha$-\ce{HgS}, \ce{Te}, and $\alpha$-quartz under similar experimental conditions. 
These splittings originate from the same underlying physical mechanism, namely, symmetry-allowed lifting of degeneracy for chiral phonons at finite wavevector.}
\addr{We emphasize that the physical origin and magnitude of the splittings are governed by three main factors: (i) the lattice structure, (ii) the excitation wavelength, and (iii) the phonon dispersion properties.
For factors (i) and (ii), the size of the Brillouin zone along the crystal $c$-axis and the excitation wavelength determine the phonon wavevector involved in Raman scattering. 
A longer $c$-axis and a shorter excitation wavelength lead to larger phonon wavevectors with respect to the size of the Brillouin zone, and thus larger splittings, since the splitting is approximately proportional to the phonon wavevector. 
For factor (iii), different $E_2$ phonon modes possess different degrees of chirality, which are reflected in the group velocities along the $c$-axis. 
These group velocities determine how rapidly the splitting increases with increasing phonon wavevector.}
\addr{From a symmetry viewpoint, the linear-in-$q$ splitting of chiral phonons is a generic consequence of lattice chirality. 
Recent theoretical work has shown that such a splitting can be described by an effective Hamiltonian with a pseudoscalar coupling constant, which is allowed only in chiral crystals and is determined by the lattice structure and force constants~\cite{tsunetsugu2026}.}

\begin{table*}
  \caption{\label{tab:compare_exp_tho}Frequencies and energy splittings of chiral phonons in {\lCS},  
  measured using circularly polarized Raman spectroscopy and compared with first-principles calculations.
  \addr{The errors in the peak splittings were estimated from the uncertainties of the Voigt-function fits.}}
  \begin{ruledtabular}
    \begin{tabular}{cccccccc}
      & \multicolumn{3}{c}{Experiment} & \multicolumn{4}{c}{Calculation} \\ 
      \cmidrule(lr){2-4} \cmidrule(lr){5-8}
      Peak & Polarization & Frequency (\unit{cm^{-1}}) & Splitting (\unit{cm^{-1}}) & \addb{CAM} & AM & Frequency (\unit{cm^{-1}}) & Splitting (\unit{cm^{-1}}) \\
      \midrule
      \multirow{2}{*}{$E_2^{(1)}$} & LR & 306.7 & \multirow{2}{*}{{$+0.3\pm0.1$}} & {$-2$} & {$-$} & 338.8 & \multirow{2}{*}{{$+0.1$}} \\
      & RL & 307.0 &  & {$+2$} & {$+$} & 338.9 &  \\
      \multirow{2}{*}{$E_2^{(2)}$} & LR & 359.2 & \multirow{2}{*}{{$-1.0\pm0.2$}} & {$-2$} & {$+$} & 377.4 & \multirow{2}{*}{{$-0.8$}} \\
      & RL & 358.2 &  & {$+2$} & {$-$} & 376.5 &  \\
      \multirow{2}{*}{$E_2^{(3)}$} & LR & 401.1 & \multirow{2}{*}{{$+0.1\pm0.1$}} & {$-2$} & {$+$} & 422.6 & \multirow{2}{*}{{$+0.5$}} \\
      & RL & 401.2 &  & {$+2$} & {$-$} & 423.1 &  \\
      \multirow{2}{*}{$E_2^{(4)}$} & LR & 456.1 & \multirow{2}{*}{{$-1.0\pm0.3$}} & {$-2$} & {$+$} & 484.2 & \multirow{2}{*}{{$-0.7$}} \\
      & RL & 455.2 &  & {$+2$} & {$-$} & 483.5 &  \\
    \end{tabular}
  \end{ruledtabular}
\end{table*}

\subsection{Calculations of phonon \addb{CAM}}
\addr{The splittings of the $E_2$ modes can be understood from the conservation of CAM.}
\addr{The concept of CAM is analogous to that of the crystal momentum, in that it is a discrete and conserved quantity determined by crystal symmetry. 
CAM is a symmetry-based quantum number defined by the eigenvalue of the rotational (or screw) symmetry operator acting on the phonon eigendisplacement, and it characterizes how the collective atomic motion transforms under crystal rotational symmetries.}
In the left-handed $\SGl$ structure, the \addb{CAM} of phonons $m$ is defined as~\cite{zhang2015,zhang2022a,kato2023a,tateishi2025}
\begin{align}
  \addb{\qty{C_6^+\middle|\frac{4\vb{c}}{6}}\vb{u}_{j}(\vb{q})=\exp[-\frac{2\pi i}{6}\qty(m+\frac{4\vb{q}\vdot\vb{c}}{2\pi})]\vb{u}_{j}(\vb{q}),}
  \label{eq:pam}
\end{align}
where $\vb{c}$ is the primitive lattice vector along the $c$-axis of the crystal, and $\vb{u}_{j}(\vb{q})$ represents the phonon eigendisplacement of the $j$-th branch at wavevector $\vb{q}$.
%
The \addb{CAM} of phonons in hexagonal crystals takes integer values $m=0,\pad \pm1,\pad \pm2,\pad 3$.

\begin{figure}
  \centering
  \includegraphics[width=\linewidth]{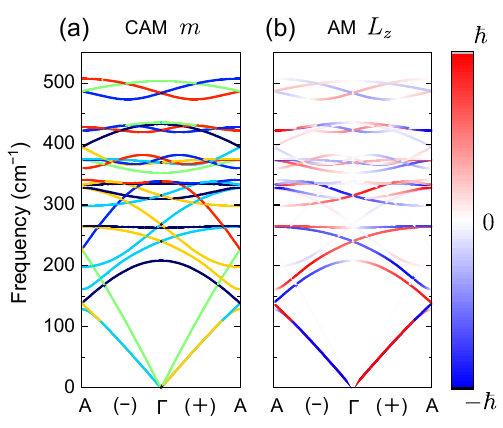}
  \caption{\label{fig:calculation}(a) \addb{CAM} $m$ and (b) the $z$ component of the angular momentum $L_z$ for phonons in {\lCS}:
  the \addb{CAM} values $m=-2,\pad -1,\pad 0,\pad +1,\pad +2$, and $3$ are shown in blue, sky blue, green, yellow, red, and black, respectively.}
\end{figure}

The CAM was numerically calculated for phonons in \ce{CrSi2} along the $\Gamma$--A line using Eq.~\eqref{eq:pam}, based on the results of the first-principles calculations.
The results for {\lCS} are shown in Fig.~\subref{fig:calculation}{a}.
Each doubly degenerate mode at the $\Gamma$ point clearly splits into two non-degenerate modes with opposite \addb{CAM} values at finite wavevectors along the $\Gamma$--A direction.
The $E_2$ modes correspond to \addb{CAM} values of $m=\pm2$ (colored red and blue).

\begin{figure}
  \centering
  \includegraphics[width=\linewidth]{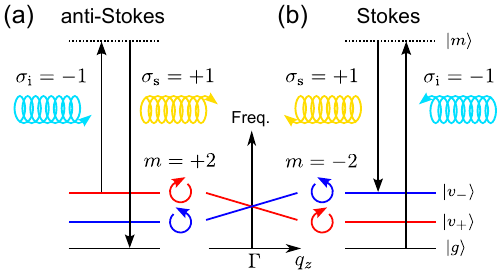}
  \caption{\label{fig:selection_rule}\addr{Schematic illustration of Raman selection rules:
  (a) anti-Stokes and (b) Stokes processes.
  $\ket{g}$ denotes the ground state, while $\ket{m}$ represents an intermediate state involved in the Raman transitions and $\ket{v_{+(-)}}$ represents an excited state with CAM $m=\pm2$ phonons.
  In this schematic, the incident and scattered photons carry CAM $\sigma_{\mathrm{i}}=-1$ and $\sigma_{\mathrm{s}}=+1$, respectively (LR configuration), 
  resulting in $m=+2$ for the Stokes process and $m=+2$ for the anti-Stokes process.}}
\end{figure}

The polarization selection rules\addb{,} or equivalently, the conservation of \addb{CAM} in the Raman process~\cite{tatsumi2018}\addb{,} lead to the following condition:
\begin{align}
  \sigma_{\mathrm{i}} - \sigma_{\mathrm{s}} \equiv \pm m \pmod{6},
\end{align}
where the plus (minus) sign corresponds to the Stokes (anti-Stokes) process, and $\sigma_{\mathrm{i(s)}}=\pm1$ denotes the \addb{CAM} of the incident (scattered) circularly polarized light.
\addr{The schematic illustration of the conservation of CAM is shown in Fig.~\ref{fig:selection_rule}.}
Thus, in the cross-circular polarization configurations, only $E_2$ phonons with $m = \pm 2$ can be excited.
Specifically, the $^1E_2$ phonon ($m = -2$) contributes to the LR configuration \addb{($\sigma_{\mathrm{i}}=-1,\ \sigma_{\mathrm{s}}=+1)$}, while the $^2E_2$ phonon ($m = +2$) contributes to the RL configuration \addb{($\sigma_{\mathrm{i}}=+1,\ \sigma_{\mathrm{s}}=-1)$} in the Stokes process.

In Raman scattering, phonons possess a small but finite wavevector because of the conservation of crystal momentum.
In the Stokes process of the present case, the phonon wavevector was approximately 1\% of the Brillouin zone size.
At positive wavevectors corresponding to the Stokes process,
phonons with \addb{CAM} of $m=+2$ exhibited higher frequencies for the $E_2^{(1)}$ and $E_2^{(3)}$ modes
but lower frequencies for the $E_2^{(2)}$ and $E_2^{(4)}$ modes.
These calculated energy splittings are consistent with the experimental results,
as listed in Table~\ref{tab:compare_exp_tho}.

\addb{In addition}, the sign of the \addb{CAM} in {\rCS} was opposite to that in {\lCS}, as shown in \addb{Appendix~\ref{app:am_right}}.
These results demonstrate that the observed splittings of the $E_2$ phonon modes originate from \addb{CAM} conservation.
At the same time, they indicate that the handedness of a \ce{CrSi2} crystal can be identified from the direction of the splittings observed in Raman scattering.
\addb{Note that} \addb{t}he handedness of the LDFZ-grown crystals determined by Raman measurements was fully consistent with that determined by XRD.

\subsection{Raman measurements of CVT-grown crystals}
\begin{figure}
  \centering 
  \includegraphics[width=0.8\linewidth]{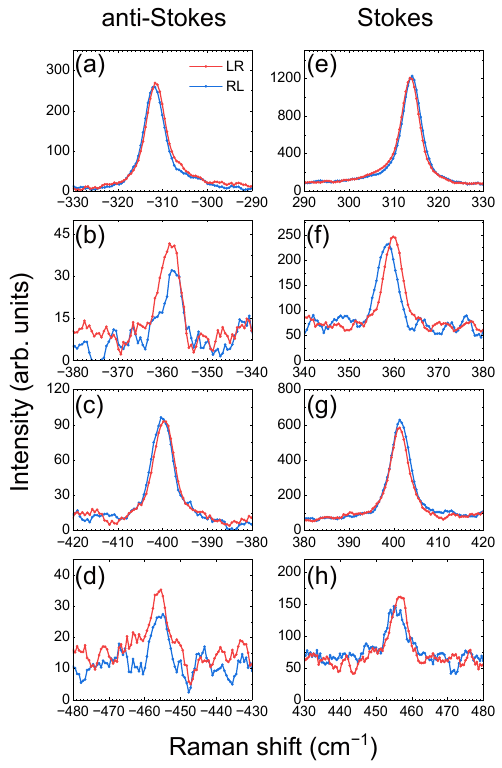}
  \caption{(a--d) Anti-Stokes and (e--h) Stokes spectra of the four $E_2$ modes of Sample~\#1 in cross-circular polarization configurations:
  (a), (e) $E_2^{(1)}$, (b), (f) $E_2^{(2)}$, (c), (g) $E_2^{(3)}$, and (d), (h) $E_2^{(4)}$ peaks,
  showing polarization-dependent energy splittings.}
  \label{fig:cvt1_spectra}
\end{figure}

For completeness, we also measured two CVT-grown crystals, which exhibited sharper Raman lines than the LDFZ-grown samples \addb{(Appendix~\ref{app:cvt_mode_assignment})}.
Stokes and anti-Stokes spectra of Sample~\#1\addb{,} \addb{identified as {\lCS} by XRD measurements,} obtained in the cross-circular polarization configurations are provided in Figs.~\subref{fig:cvt1_spectra}{a}--\subref{fig:cvt1_spectra}{h}.
All four $E_2$ phonon modes showed clear differences in Raman shift between the \addb{LR and RL} spectra.
The direction of the splittings was consistent with that of the {\lCS} grown by the LDFZ method.
In contrast, the anti-Stokes spectra exhibited splittings in the opposite direction to the Stokes spectra, as expected from the reversal of the \addb{CAM} at negative wavevectors [Fig.~\subref{fig:calculation}{a}].
These results further confirm that the observed $E_2$ phonon splittings strictly follow the selection rules derived from \addb{CAM} conservation in {\lCS}, consistent with the handedness determined by XRD.

\begin{figure}
  \centering 
  \includegraphics[width=0.8\linewidth]{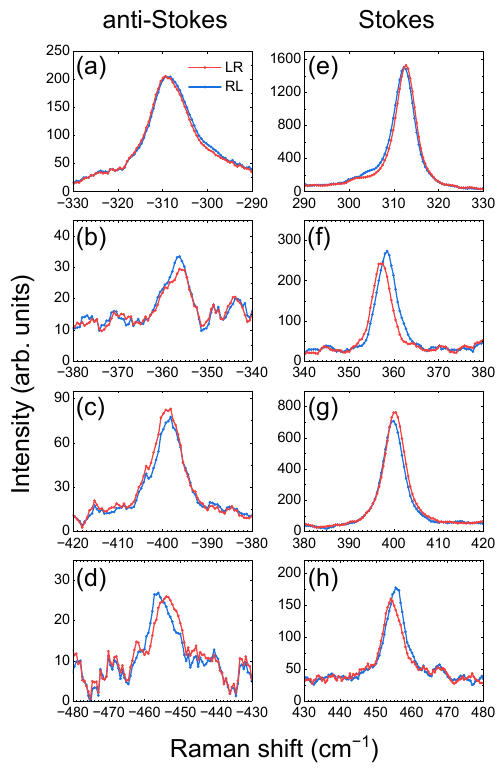}
  \caption{(a--d) Anti-Stokes and (e--h) Stokes spectra of the four $E_2$ modes of Sample~\#2 in cross-circular polarization configurations:
  (a), (e) $E_2^{(1)}$, (b), (f) $E_2^{(2)}$, (c), (g) $E_2^{(3)}$, and (d), (h) $E_2^{(4)}$ peaks,
  showing polarization-dependent energy splittings.}
  \label{fig:cvt2_spectra}
\end{figure}

Similar splittings of the $E_2$ modes were also observed in Sample~\#2 [Figs.~\subref{fig:cvt2_spectra}{a}--\subref{fig:cvt2_spectra}{h}].
Importantly, the relation between the \addb{LR and RL} spectra was consistent with the {\rCS} case summarized in Table~\ref{tab:compare_sample1_2}.
Thus, the handedness of Sample~\#2, which could not be determined by XRD, can be identified as right-handed \addb{by Raman spectroscopy}.

\begin{table*}
  \caption{\label{tab:compare_sample1_2}Frequencies and energy splittings of chiral phonons in CVT-grown \ce{CrSi2} crystals,  
  measured in the cross-circular polarization configurations.
  \addr{The errors in the peak splittings were estimated from the uncertainties of the Voigt-function fits.}}
  \begin{ruledtabular}
    \begin{tabular}{cccccc}
      & & \multicolumn{2}{c}{Sample~\#1} & \multicolumn{2}{c}{Sample~\#2} \\ 
      \cmidrule(lr){3-4} \cmidrule(lr){5-6}
      Peak & Polarization & Frequency (\unit{cm^{-1}}) & Splitting (\unit{cm^{-1}}) & Frequency (\unit{cm^{-1}}) & Splitting (\unit{cm^{-1}}) \\
      \midrule
      \multirow{2}{*}{$E_2^{(1)}$} & LR & 313.6 & \multirow{2}{*}{{$+0.3\pm0.1$}} & 312.6 & \multirow{2}{*}{{$-0.4\pm0.1$}} \\
      & RL & 314.0 & & 312.2 &  \\
      \multirow{2}{*}{$E_2^{(2)}$} & LR & 359.9 & \multirow{2}{*}{{$-1.4\pm0.1$}} & 357.3 & \multirow{2}{*}{{$+1.1\pm0.1$}} \\
      & RL & 358.4 & & 358.4 &  \\
      \multirow{2}{*}{$E_2^{(3)}$} & LR & 401.2 & \multirow{2}{*}{{$+0.2\pm0.1$}} & 400.2 & \multirow{2}{*}{{$-0.4\pm0.1$}} \\
      & RL & 401.5 & & 399.9 &  \\
      \multirow{2}{*}{$E_2^{(4)}$} & LR & 456.5 & \multirow{2}{*}{{$-1.0\pm0.3$}} & 454.6 & \multirow{2}{*}{{$+0.9\pm0.2$}} \\
      & RL & 455.5 & & 455.5 &  \\
    \end{tabular}
  \end{ruledtabular}
\end{table*}

\subsection{CAM and angular momentum of phonons}
\addr{At small but finite wavevector $\vb{q}$, the lifting of degeneracy allows phonon modes with opposite CAM to split in energy, and the corresponding branches acquire opposite angular momentum. 
Importantly, CAM and phonon angular momentum are closely related but fundamentally distinct quantities~\cite{kato2023a,tateishi2025,zhang2026}: 
the former classifies phonons according to the crystal symmetry, whereas the latter measures the real-space atomic rotation.}

\addr{The transformation properties of phonons for the crystal symmetry are associated with the irreducible representations of the point group $6$ ($C_6$). 
As listed in Table~\ref{tab:character_pam}, each phonon CAM can be directly related to the irreducible representation of the corresponding vibrational mode.}
\begin{table}
  \caption{\label{tab:character_pam}Character table of the point group $6$ ($C_6$) and corresponding \addb{CAM} $m$: here, $\omega = \exp(2\pi i/3)$.}
  \begin{ruledtabular}
    \begin{tabular}{cccccccc}
      Irreps. & {$E$} & {$C_6^+$} & {$C_3^+$} & {$C_2$} & {$C_3^-$} & {$C_6^-$} & {\addb{CAM}} \\
      \midrule
      $A$ & 1 & 1 & 1 & 1 & 1 & 1 & 0 \\
      $B$ & 1 & {$-1$} & 1 & {$-1$} & 1 & {$-1$} & 3 \\
      $^1E_2$ & 1 & {$\omega$} & {$\omega^2$} & 1 & {$\omega$} & {$\omega^2$} & {$-2$} \\
      $^2E_2$ & 1 & {$\omega^2$} & {$\omega$} & 1 & {$\omega^2$} & {$\omega$} & {$+2$} \\
      $^2E_1$ & 1 & {$-\omega^2$} & {$\omega$} & {$-1$} & {$\omega^2$} & {$-\omega$} & {$-1$} \\
      $^1E_1$ & 1 & {$-\omega$} & {$\omega^2$} & {$-1$} & {$\omega$} & {$-\omega^2$} & {$+1$} \\
    \end{tabular}
  \end{ruledtabular}
\end{table}

\addr{In contrast, the phonon angular momentum originates from the microscopic rotational motion of individual atoms and quantifies the actual mechanical rotation associated with a phonon mode. 
While CAM is a discrete quantum number that remains constant within a given phonon branch, angular momentum is a continuous quantity that depends on the phonon wavevector $\vb{q}$.}
\addb{W}e calculated the phonon angular momentum \addb{(Appendix~\ref{app:eq_am} for details)} to verify whether the phonons observed in the Raman spectra correspond to chiral phonons\addb{.} 
As shown in Fig.~\subref{fig:calculation}{b}, the $E_2$ modes exhibited finite angular momenta along their wavevectors. 
Therefore, the $E_2$ phonons observed in the cross-circular polarization configurations can be identified as chiral phonons~\cite{ishito2023a}.

\section{Conclusion}
In summary, circularly polarized Raman measurements were performed on the sixfold chiral semiconductor \ce{CrSi2}.
Under normal incidence along the crystal $c$-axis, energy splittings between the two cross-circular polarization configurations were observed in four $E_2$ phonon modes.
These splittings are consistent with the conservation of phonon \addb{CAM}.
First-principles calculations of phonon angular momentum further confirmed that the observed Raman features originate from chiral phonons.
The results establish circularly polarized Raman scattering as a powerful probe of chiral phonons and crystal handedness in hexagonal chiral systems, expanding the exploration beyond trigonal materials.
\addr{Moreover, the splitting of doubly degenerate $E_1$ or $E_2$ modes at the $\Gamma$ point into nondegenerate branches suggests a potential route to exploring topological phonon properties.}


\section{Acknowledgments}
G.K. was supported by JST SPRING (Grant No. JPMJSP2180).
Y.K. was supported by JSPS KAKENHI (Grant No. JP23H01870).
Y.T. was supported by JSPS KAKENHI (Grant Nos. JP22H01944, JP23H01870 and JP23H00091) and 
JSPS International Joint Research Program (JRP-LEAD with UKRI) (Grant No. JPJSJRP20241710).
T.S. was supported in part by JSPS KAKENHI (Grant Nos. JP21H01032 and JP22H01154), 
MEXT X-NICS (Grant No. JPJ011438),  
and JST CREST (Grant No. JPMJCR24R5).
Y.T. and T.S. were supported by NINS OML Project (Grant No. OML012301) \addb{and JST ERATO (Grant No. JPMJER2503)}.

\appendix

\setcounter{figure}{0}
\renewcommand{\thefigure}{A\arabic{figure}}
\setcounter{table}{0}
\renewcommand{\thetable}{A\arabic{table}}
\section{\label{app:selection_rule}Raman selection rules}
\subsection{Derivation of Raman tensors}
The Stokes intensity of Raman scattering is proportional to $I\propto|\vb{e}_{\mathrm{s}}^\dagger R\vb{e}_{\mathrm{i}}|^2$,
where $\vb{e}_{\mathrm{i(s)}}$ denotes the polarization vector of the incident (scattered) light, and $R$ is the Raman tensor.
The selection rules are determined by the form of $R$, which in turn is dictated by the crystal symmetry.

Because the phonon wavevector involved in Raman scattering is small compared with the size of the Brillouin zone,
Raman tensors are usually derived from the point group at the $\Gamma$ point.
However, for finite $\vb{q}$, which plays a crucial role in the present case,
the symmetry of the phonons and their Raman tensors is governed by the little co-group (the subgroup of the crystal point group that leaves $\vb{q}$ invariant).

For the present case of the hexagonal point group $622$ ($D_6$), with the phonon wavevector along the $c$-axis,
the relevant symmetry operations reduce to those of the point group $6$ ($C_6$).
The corresponding character table is listed in Table~{II} of the main text.
Raman tensors adapted to the irreducible representations of point group $6$ can be obtained using the projection operator, defined as~\cite{dresselhaus2007}
\begin{align}
  P_{ij}^{r}=\frac{d_{r}}{g}\sum_{R\in G}D_{ij}^{r*}(R) R,
\end{align}
where $R$ denotes a symmetry operation of the group $G$, $g$ is the order of the group, $d_r$ is the dimension of the irreducible representation $r$,
and $D^{r}(R)$ is its representation matrix.
Applying this operator to a trial matrix yields the Raman tensors for
\begin{gather}
  R^{A}=
  \mqty(a & 0 & 0 \\
  0 & a & 0 \\
  0 & 0 & b),\\
  R^{^1E_2}=
  c\mqty(1 & i & 0 \\
  i & -1 & 0 \\
  0 & 0 & 0),\quad
  R^{^2E_2}=
  d\mqty(1 & -i & 0 \\
  -i & -1 & 0 \\
  0 & 0 & 0).
\end{gather}
Here, only the symmetric components are shown.
The superscripts of the Raman tensors indicate the phonon modes excited in the Stokes process.

\subsection{Selection rules for the circular polarization configurations}
Circular polarization vectors are defined as \addb{$\vb{e}_{\mathrm{L}}=\transp{\qty(1/\sqrt{2},\pad -i/\sqrt{2},\pad 0)}$} and \addb{$\vb{e}_{\mathrm{R}}=\transp{\qty(1/\sqrt{2},\pad i/\sqrt{2},\pad 0)}$}.
Using these definitions, the Raman scattering intensities of Stokes light for the $A$ and $E_2$ modes can be evaluated as
\begin{align}
  I_{\mathrm{parallel}}^{A}\propto a^2,\quad I_{\mathrm{cross}}^{A}=0,
\end{align}
for the $A$ mode, and
\begin{align}
  I_{\mathrm{LR}}^{^1E_2}&\propto c^2,\quad I_{\mathrm{RL}}^{^1E_2}=I_{\mathrm{parallel}}^{^1E_2}=0, \\
  I_{\mathrm{RL}}^{^2E_2}&\propto d^2,\quad I_{\mathrm{LR}}^{^2E_2}=I_{\mathrm{parallel}}^{^2E_2}=0,
\end{align}
for the $E_2$ modes.
Thus, the $A$ mode appears only in the parallel-circular \addb{polarization} configuration\addb{s},
the $^1E_2$ ($m=-2$) mode only in the LR configuration,
and the $^2E_2$ ($m=+2$) mode only in the RL configuration.
The obtained selection rules are consistent with the conservation law of \addb{CAM}.

Formally, the intensities of anti-Stokes light can be obtained by replacing the Raman tensors as $R \to R^*$,
since the Stokes and anti-Stokes processes are related by time-reversal symmetry \addb{in the limit of zero-phonon frequency}.
Thus, the intensities can be expressed as
\begin{align}
  I_{\mathrm{parallel, AS}}^{A}&\propto a^2,\quad I_{\mathrm{cross, AS}}^{A}=0, \\
  I_{\mathrm{LR, AS}}^{^2E_2}&\propto d^2,\quad I_{\mathrm{RL, AS}}^{^1E_2}=I_{\mathrm{parallel}}^{^1E_2}=0, \\
  I_{\mathrm{RL, AS}}^{^1E_2}&\propto c^2,\quad I_{\mathrm{LR, AS}}^{^2E_2}=I_{\mathrm{parallel}}^{^2E_2}=0,
\end{align}
indicating that the $^1E_2$ ($m=-2$) modes contribute only to the RL configuration,
whereas the $^2E_2$ ($m=+2$) modes contribute only to the LR configuration.
Therefore, the conservation law of \addb{CAM} is preserved in the anti-Stokes process as well.

\setcounter{figure}{0}
\renewcommand{\thefigure}{B\arabic{figure}}
\setcounter{table}{0}
\renewcommand{\thetable}{B\arabic{table}}
\section{\label{app:am_right}Calculation of phonons in the right-handed structure}
\begin{figure}[H]
  \centering
  \includegraphics[width=\linewidth]{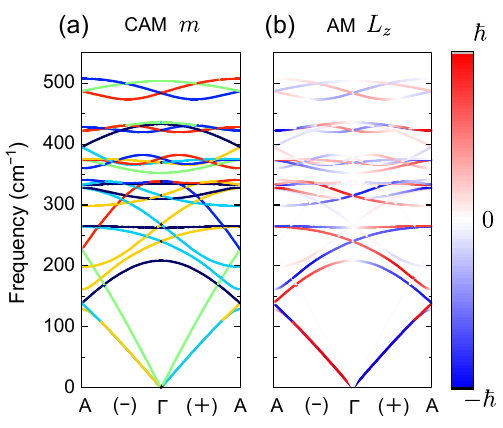}
  \caption{\label{fig:r_calculation}(a) \addb{CAM} $m$ and (b) the $z$ component of the angular momentum $L_z$ for phonons in {\rCS}:
  the \addb{CAM} values $m=-2,\pad -1,\pad 0,\pad +1,\pad +2$, and $3$ are shown in blue, sky blue, green, yellow, red, and black, respectively.}
\end{figure}

In the right-handed $\SGr$ structure, the \addb{CAM} of the phonons $m$ is defined as~\cite{zhang2015,zhang2022a,kato2023a,tateishi2025}
\begin{align}
  \addb{\qty{C_6^+\middle|\frac{2\vb{c}}{6}}\vb{u}_{j}(\vb{q})=\exp[-\frac{2\pi i}{6}\qty(m+\frac{2\vb{q}\vdot\vb{c}}{2\pi})]\vb{u}_{j}(\vb{q}).}
\end{align} 
Figures~\subref{fig:r_calculation}{a} and \subref{fig:r_calculation}{b} show the calculated results for the \addb{CAM} and the phonon angular momentum in {\rCS}.
The signs of both quantities are reversed compared with those of the {\lCS}, which is consistent with the opposite crystal chirality.

\setcounter{figure}{0}
\renewcommand{\thefigure}{C\arabic{figure}}
\setcounter{table}{0}
\renewcommand{\thetable}{C\arabic{table}}
\section{\label{app:cvt_mode_assignment}Mode assignments of the chemical vapor transport crystal}
\begin{figure}[H]
  \centering
  \includegraphics[width=\linewidth]{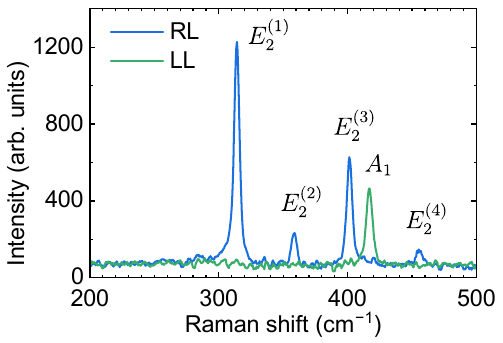}
  \caption{\label{fig:cvt1_mode_assignment}Phonon mode assignments for Sample~\#1 grown by the CVT method.}
\end{figure}

Figure~\ref{fig:cvt1_mode_assignment} shows the phonon mode assignments of Sample~\#1.
Compared with the LDFZ-grown crystals discussed in the main text, the CVT-grown crystal exhibited sharper Raman lines, indicating higher crystallinity.

\setcounter{figure}{0}
\renewcommand{\thefigure}{D\arabic{figure}}
\setcounter{table}{0}
\renewcommand{\thetable}{D\arabic{table}}
\section{\label{app:eq_am}Calculations of phonon angular momentum}
The phonon angular momentum is calculated using the following expression~\cite{zhang2014}:
\begin{align}
  \vb{L}_{j}(\vb{q}) = \hbar \vb*{\epsilon}_{j}^{\dagger}(\vb{q})\vb{M}\vb*{\epsilon}_{j}(\vb{q}),
\end{align}
where $\vb*{\epsilon}_{j}(\vb{q})$ is the normalized eigenvector of the dynamical matrix. 
For the $z$ component, the matrix $\vb{M}$ is given by
\begin{align}
  M_z=I_{N\times N}\otimes\mqty(0 & -i & 0 \\ i & 0 & 0 \\ 0 & 0 & 0).
\end{align}


\bibliography{CrSi2_arXiv.bbl}

\end{document}

%% file: revpackage.tex
\usepackage[pdftex]{graphicx}
\usepackage{here}

\usepackage{booktabs} 
\usepackage{multirow}

\usepackage{bm}
\usepackage{xcolor}
    
\usepackage{amsmath, amssymb, amsfonts}

\usepackage[colorlinks=true,allcolors=blue]{hyperref}

\usepackage{siunitx} 
\NewCommandCopy\sqty\qty 
\NewDocumentCommand\Sqty{smm}{%
    \IfBooleanTF{#1}%
    {\sqty[quantity-product={}]{#2}{#3}}%
    {\sqty[quantity-product=~]{#2}{#3}}%
}
\usepackage[italicdiff]{physics} 
\usepackage[version=4]{mhchem} 

\usepackage{cleveref}
\crefname{equation}{Eq.}{Eqs.} 
\crefname{figure}{Fig.}{Figs.}
\Crefname{figure}{Figure}{Figures}

\newcommand{\transp}[1]{#1^\top} 
\NewDocumentCommand\qtxt{sm}{%
    \IfBooleanTF{#1}%
    {~\textrm{#2}}%
    {\quad\textrm{#2}}%
}
\newcommand{\pad}{\,}

\NewDocumentCommand{\subref}{mm}{\hyperref[#1]{\ref{#1}(#2)}}
\graphicspath{%
{./}%
}

\newcommand{\SGl}{P6_422}
\newcommand{\SGr}{P6_222}
\newcommand{\lCS}{L-\ce{CrSi2}}
\newcommand{\rCS}{R-\ce{CrSi2}}

\usepackage[normalem]{ulem}
\DeclareRobustCommand{\eraser}{\bgroup\markoverwith{\textcolor{red}{\rule[.5ex]{2pt}{0.4pt}}}\ULon}
\DeclareRobustCommand{\eraseb}{\bgroup\markoverwith{\textcolor{blue}{\rule[.5ex]{2pt}{0.4pt}}}\ULon}
\newcommand{\addr}[1]{\textcolor{red}{#1}}
\newcommand{\addb}[1]{\textcolor{blue}{#1}}